\documentclass{aipproc}
\layoutstyle{8x11single}
\usepackage{rotate}
\usepackage{epsfig}
\def\bge{\begin{equation}}
\def\ene{\end{equation}}
\newcommand{\nc}{\newcommand}
\nc{\non}{\nonumber}
\def\be{\begin{equation}}
\def\ee{\end{equation}}
\def\bga{\begin{eqnarray}}
\def\ena{\end{eqnarray}}
\def\eea{\end{eqnarray}}
\def\bg{\begin{eqnarray}}
\def\en{\end{eqnarray}}

\def\ra{\rightarrow}

\def\hbar{\not\!h}

\begin{document}

\title 
      [Boundary of Nuclear Physics and QCD]
      {Boundary of Nuclear Physics and QCD}

\thanks{Invited presentation at the International Conference on
Nuclear Physics (INPC2001): Berkeley, July 30 to August 3, 2001 \\
University of Adelaide preprint: ADP-01-32/T464}
\author{A.~W.~Thomas}{
  address={Department of Physics and Mathematical Physics and \\
  Special Research Centre for the Subatomic Structure of Matter \\
  The University of Adelaide, Adelaide SA 5005, Australia},
  email={athomas@physics.adelaide.edu.au}
}


\copyrightyear  {2001}

\begin{abstract}
Recent progress in lattice QCD, combined with the imminent advent of a
new generation of dedicated supercomputers and advances in chiral
extrapolation mean that the next few years will bring quite novel insights into
hadron structure.  We review some of the recent highlights in this field,
the questions which might be addressed and the experiments which may be
expected to stretch that understanding to its limits.  Only with a sound
understanding of hadron structure can one hope to explore the fundamental
issue of how that structure may change at finite density (or
temperature).
We explore potential future insights from lattice QCD into the
phenomenon of nuclear saturation and a very important hint from recent
data of a change in the structure of a bound nucleon.
\end{abstract}

\date{\today}

\maketitle

\section{Introduction}
Modern nuclear physics presents a wonderfully exciting and diverse set of
challenges.  Perhaps the most fundamental of these is the challenge to 
understand nuclear phenomena in terms of the underlying theory of the 
strong interaction - Quantum Chromodynamics or QCD.  At this meeting we saw 
some excellent discussions of QCD at high temperature ($T$) and density 
($\rho$), especially in the context of relativistic heavy ion collisions, 
the quark-gluon plasma and beyond.  We concentrate on the lower density 
regime which is generally termed hadronic physics.  This is the other sector 
of nuclear physics which can naturally be addressed in terms of QCD.

Hadronic physics is currently at an extremely exciting stage of 
development.  New experimental capabilities 
at laboratories such as JLab, Mainz 
and MIT-Bates are extending our knowledge of nucleon form factors free 
and bound (as well as transition form factors) into new kinematic domains 
and with unheard of precision.  We are beginning to see the development
of phenomenologically meaningful, covariant models of hadron structure 
and these can be extended to incorporate chiral symmetry.  With the 
development of clever improved actions, faster computers and better treatments
of chiral symmetry, lattice QCD is soon to deliver on its promise of deep
new insights into hadron structure.  Indeed, as we shall describe, in 
combination with carefully controlled chiral extrapolation one can expect 
to calculate accurate properties of the low mass baryons within a few years.
One can also hope to rigorously address some key physics issues in hadron
spectroscopy.

In the decade since Brown and Rho focused attention on the possible 
change of hadron properties in-medium, the topic has generated enormous 
theoretical and experimental interest.  We briefly outline the role of changes 
of nucleon internal structure within relativistic mean field theory.  We also
suggest how lattice input might feed directly into such calculations. 
{}Finally, we report on what has the potential to be a very important 
development in this field, namely the recent determination of $G_E/G_M$ for a 
proton bound in $^4$He.  The relative insensitivity of the
analysis to various theoretical
corrections, combined with the apparent deviation from the free
$G_E/G_M$ ratio, will stimulate a great deal of interest.  In the context of 
understanding nuclear structure in terms of QCD, it is the first firm 
evidence for the change in the structure of a bound nucleon which must be 
there.

In the next section we make some general remarks about QCD, its bizarre
properties and the nature of the QCD vacuum.  After describing recent
advances in linking lattice QCD to covariant models of hadron structure
we summarise the crucial issue of chiral extrapolation.  The recent
progress in this area, for which we present several examples, underpins
the exciting prospect that we may be able to calculate accurate hadronic
properties at the physical quark mass in just a few years.  We close the
discussion with an outlook of the recent progress in hadron
spectroscopy.  The third section deals with the in-medium 
properties of hadrons, while 
the final section contains some closing remarks.
\begin{figure}[tbp]
\centering{\
\epsfig{figure=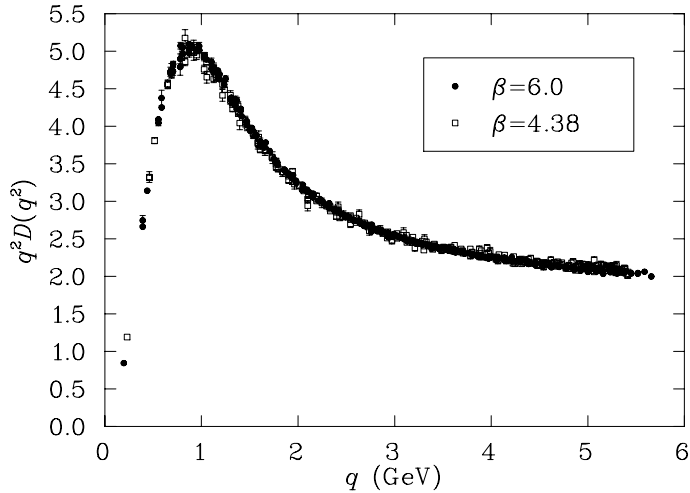,angle=90,width=9cm}
\caption{
Non-perturbative behaviour of the gluon propagator (times $q^2$)
in Landau gauge,
calculated from lattice QCD -- from Ref. \protect\cite{Bonnet:2001uh}. }
\label{fig:AllProps}}
\end{figure}
%

\section{Lattice QCD and Hadron Structure}
The bizarre properties of QCD are well known.  It exhibits asymptotic
freedom at short distances while it is confining in the long distance
regime.  At least qualitatively one can understand confinement in terms
of non-trivial QCD vacuum structure -- essentially a dual
superconductor \cite{Toki:2000er}. 
This confines the colour electric fields (dual to the
magnetic fields in the case of a normal superconductor) between quarks
into a flux tube of approximately constant area.  Hence the energy to
separate two quarks grows linearly with separation.  To put the forces
being discussed into context, it is worth noting that the rate of growth
of the energy, the string tension, is approximately 1 GeV/fm 
($\sigma =$ (440 MeV)$^2$).
This corresponds to a constant, confining force of order 10
tonnes --  a force characteristic of trucks acting 
between objects less than $10^{-18}$m
in size!

The non-trivial nature of the QCD vacuum is illustrated by the fact that
it contains both quark and gluon condensates \cite{Thomas:2001kw}.  For a 
purely gluonic version of QCD the vacuum energy density, $\epsilon_{\rm
vac}$, is \cite{Novikov:1981xj}:
\be
\epsilon_{\rm vac} = - \frac{9}{32} \langle 0 | \frac{\alpha_s}{\pi}
G^2 | 0 \rangle = - 0.5 {\rm GeV/fm}^3.
\label{eq:1}
\ee
In comparison with phenomenological estimates of the
energy difference between the
perturbative and non-perturbative vacuum states, such as $B$ in the MIT
bag model, this is an order of magnitude larger. Thus, either the popular
idea of the perturbative vacuum being fully restored inside a hadron is
incorrect or the situation is rather more complicated than commonly
assumed.

These gluon fields in vacuum also show important topological structure,
such as the famous instantons \cite{'tHooft:1976fv}. 
This topology is believed to be
connected with chiral symmetry breaking, which we will consider soon.
{}For the present we note that current lattice simulations allow us to
actually display this topological structure, to correlate it with
regions of high action density and to examine its development in
Euclidean time.  It is impossible to display such information here, but
the interested reader is referred to various animations, generated by
Leinweber and collaborators, on the CSSM web
pages \cite{CSSM} (see also Ref. \cite{Bonnet:2001rc}).
At first glance it is clear that the
topological structures are neither spherical nor weakly interacting, but
the benefits of these visualizations are just beginning to be
appreciated.

Covariant models of hadron structure are very much in their infancy.
Nevertheless, substantial progress has been made in understanding the
structure of the low-lying pseudoscalar and vector mesons within a
phenomenological implementation of the 
Dyson-Schwinger equations \cite{Tandy:1996sq,Roberts:1994dr}. In
addition, there have been some promising developments in the baryon
sector based on the Faddeev equations \cite{Alkofer:2001ne}. 
Until now the phenomenological
input has been chosen to reproduce some limited set of experimental data
and then applied to other problems. However, the sophistication of
modern lattice gauge theory is such that one can now begin to check key
parts of these covariant calculations against lattice simulations.

The natural starting point for comparisons between covariant
calculations and lattice simulations are the quark and gluon
propagators. The long term aim is to refine the model 
building process using QCD itself. Of course, intermediate steps such as
the quark and gluon propagators are not physical and one must
specifically fix the gauge in order to make a meaningful comparison. The
gauge most commonly used is Landau gauge and techniques have been
developed to fix lattice quantities in this gauge. Figure 1 shows the
result for the non-trivial momentum dependence of the gluon propagator
(times $q^2$) \cite{Bonnet:2001uh},
$q^2 D(q^2)$, which should go to a constant at large $q^2$ (up to
perturbative QCD logs). From Fig. 1
we see that the lattice simulation shows that the gluon
propagator is clearly {\it non-perturbative} for $q^2 < 4$GeV$^2$.
Even more interesting from the point of view of model building is the
fact that the gluon propagator is not enhanced as $q^2 \ra 0$.
While this agrees with some recent Schwinger-Dyson studies of QCD
\cite{Alkofer:2000mz}, it is in disagreement with at least a naive
interpretation
of a great deal of phenomenological work related to dynamical chiral
symmetry breaking within that formalism 
\cite{Hawes:1998cw}. Clearly this sort of interplay between phenomenological
models and QCD itself has just begun and we have a great deal to learn
from it.

\begin{figure}[tbp]
\centering{\
\epsfig{figure=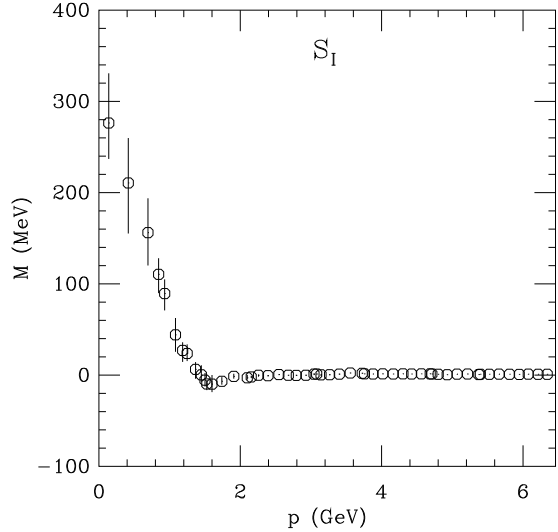, height=7cm}
\caption{
Non-perturbative behaviour of the quark mass function, $M(p^2)$,
extrapolated linearly to the chiral limit, on the basis of 
lattice QCD simulations -- from Ref. \protect\cite{Skullerud:2001aw}. }
\label{fig:QuarkProp}}
\end{figure}
Again with Landau gauge fixing, there have been some preliminary studies
of the quark propagator in QCD. For Euclidean $p^2$ one can write the quark
propagator as:
\be
S_E(p) = \frac{Z(p^2)}{i \gamma_\mu p_\mu + M(p^2)}.
\label{eq:4}
\ee
The lattice simulations, which have so far been carried out with
relatively large current quark masses, show a clear enhancement in the
infrared \cite{aoki,Skullerud:2001aw}. For example, for
a current quark mass of order 110 MeV, the
simulations suggest $M(0) \sim 400$ MeV, decreasing to around 300 MeV in
the chiral limit. We illustrate the mass function, $M(p^2)$, in the
chiral limit, as calculated by Leinweber et al. \cite{Skullerud:2001aw}, 
in Fig.~\ref{fig:QuarkProp}. 
The enhancement in the infrared region, leading to a quark 
effective mass of order 300 MeV, is clearly 
consistent with the general idea of
the constituent quark model. Indeed this result provides a firm
theoretical foundation for the concept within QCD. Of course, it also
indicates where the concept breaks down and it is clear that in
processes involving significant momentum transfer it will be necessary
to go beyond the simple idea of a fixed mass. The similarity of the
mass function, $M(p^2)$, to that found in Schwinger-Dyson studies
suggests that the latter may be a promising phenomenological
extension of the constituent quark idea.

One of the crucial tests of QCD itself is the quest for hadrons in which
gluons play a genuine structural role. For example, the
experimental discovery of exotic mesons, where the quantum numbers
cannot be associated with a $q \bar q$ pair alone, 
would be a vital step towards a full
understanding of QCD. This explains the excitement over the
announcement,
from E852 at Brookhaven National Lab \cite{Ivanov:2001rv},
of three candidates for hybrid mesons with quantum numbers $J^{PC} =
1^{-+}$. The $\pi '(1370)$ was seen in the $\pi \eta$ and $\pi \eta '$
channels, the $\pi '(1640)$ in $\pi \eta ', \rho \pi$ and $f_1 \pi$ and
the $\pi '(2000)$ in $a_1 \eta$. These masses are somewhat lower than
the values usually reported in lattice simulations, although for the
moment the latter tend to be based on quenched QCD 
\cite{Lacock:1999be,Bernard:1997ib}.
While the interpretation of the BNL data should
become clearer over the next few years, the announcement lends even
greater urgency to the calls for a future HALL D program at Jefferson
Lab \cite{Alex}.

An even more dramatic prediction of QCD than exotic states is, of
course, the possibility of physical particles containing {\em only} glue
-- the glueballs. Lattice simulations suggest that the lowest mass state
of pure glue would be the $0^{++}$ with a mass of $1611 \pm 30 \pm 160$ MeV
\cite{Michael:2001qz}. Experimental searches fave so far found a number of
scalar glueball candidates in the mass region 
1300 to 1800 MeV \cite{Bugg:2001ic}. However,
the interpretation of the data is badly effected by the fact that in
{\em real} QCD, with light quarks, no physical state will be pure glue
-- rather the best one can hope for is an unstable state with only a
small $q \bar q$ component for some (unknown) dynamical reason. We note
that the channel coupling effects induced by decay channels such as $\pi
\pi$ and $K \bar K$ are also quite controversial from the theoretical
point of view. There is clearly room for a great deal of experimental
and theoretical work in this field in the future. 

\subsection{Chiral Symmetry in the Context of Lattice QCD}
\begin{figure}[htb]
\centering{\
\epsfig{file=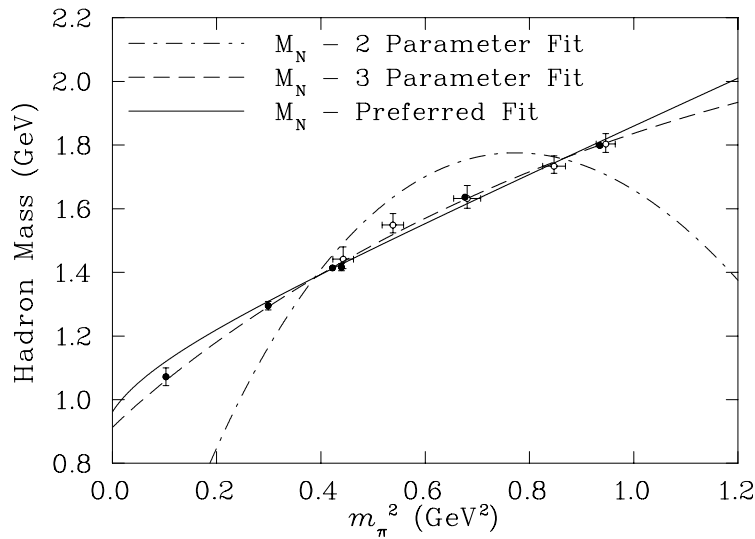, angle=90, height=7cm}
\caption{A comparison between phenomenological fitting functions for
the mass of the nucleon -- from Ref. \protect\cite{Leinweber:2000ig}.
The two parameter fit corresponds to using
Eq.(\ref{eq:5}) with $\gamma$ set equal to
the value known from $\chi$PT.  The three
parameter fit corresponds to letting $\gamma $
vary as an unconstrained fit parameter. The solid line is the
two parameter fit based on
the functional form of Eq.(\ref{eq:6}).
\label{fig:FIG3}}}
\end{figure}
The essential problem in performing calculations at realistic quark
masses (of order 5 MeV) is the approximate chiral symmetry of QCD.
Goldstone's theorem tells us that chiral symmetry is dynamically broken
and that the non-perturbative vacuum is highly non-trivial, with
massless Goldstone bosons in the limit 
$\bar m \rightarrow 0$ \cite{Thomas:2001kw}.  For finite
quark mass these
bosons are the three charge states of the pion with a
mass $m_\pi^2 \propto \bar m$.
Although this result strictly holds only for $m_\pi^2$ 
near zero (the Mann-Oakes-Renner relation),
lattice simulations show it is a good
approximation for $m_\pi^2$ up to 0.8 GeV$^2$ and we shall use
$m_\pi^2$ as a measure of the
deviation from the chiral limit.

{}From the point of view of lattice simulations with dynamical quarks
(i.e. unquenched) the essential difficulty is that the time taken goes as
$\bar m^{-3}$, or worse \cite{Lippert:2000zq}. 
The state-of-the-art for hadron masses is
$\bar m$ above 60 MeV, although
there is a preliminary result from CP-PACS at about 40 MeV
\cite{Aoki:1999ff}. In general the quark
masses for which simulations currently exist are at masses a factor of
8-20 too high.  This means that an increase of computing power to
several hundred tera-flops is needed if one is to calculate realistic hadron
properties.  Even with the current remarkable rate of increase this will
take a long time.

{}Faced with such a serious difficulty physicists (like all other people
facing a tough challenge) fall into two classes:

(A) Those who believe that ``the cup is half empty'':

In this case the emphasis is on the uncertainties associated with having
to make big extrapolations to the chiral limit.

(B) Those who believe that ``the cup is half full'':

In this case we realize that the lattice data obtained so far represents
a wealth of information on the properties of hadrons within QCD itself
over a range of quark masses.  Just as the study of QCD as a function of
$N_c$ has taught us a great deal, so the behaviour as a function of
$\bar m$ can
give us great insight into hadronic physics and guide our model
building.
{}Furthermore, as a bonus, approach (B) leads us to resolve most of the
difficulty identified under (A)!

In light of the brief space available to outline a great deal of
evidence we first summarise the conclusions which emerge from the work
of the past three years.  We then present several illustrations
of the reasoning which led to these conclusions.

{\bf Summary:}
\begin{itemize}

\item In the region of quark masses $\bar m > 60$ MeV or so ($m_\pi$
greater than typically 400-500 MeV)
hadron properties are smooth, slowly varying
functions of something like a constituent quark mass, $M \sim M_0 + c
\bar m$ (with $c \sim 1$).

\item Indeed, $M_N \sim 3 M, M_{\rho, \omega} \sim
2 M$ and magnetic moments behave like $1/M$.

\item As $\bar m$ decreases below 60 MeV or so, chiral symmetry leads to
rapid, non-analytic variation, with: \\ $\delta M_N \sim {\bar m}^{3/2},
\\
\delta \mu_H \sim {\bar m}^{1/2}$, \\ $\delta <r^2>_{\rm ch} \sim \ln
\bar m$ and \\ moments of non-singlet parton distributions 
$\sim m_\pi^2 \ln m_\pi$.

\item Chiral quark models, like the cloudy bag model (CBM) \cite{CBM},
provide a natural explanation of 
this transition. The scale is basically set by the inverse size of the
pion source -- the inverse of the bag radius in the CBM.

\item When the pion Compton wavelength is smaller than the size
of the composite source chiral loops are strongly suppressed. On the
other hand, as soon as the pion Compton wavelength is larger than the
source one begins to see rapid, non-analytic chiral corrections.
\end{itemize}

The nett result of this discovery is that one has control over the
chiral extrapolation of hadron properties provided one can get data at
pion masses of order 200--300 MeV. This seems feasible with the next
generation of supercomputers which should be available within 2--3 years
and which will have speeds in excess of 10 tera-flops \cite{LHPC}.
This is an extremely exciting possibility in that it will bring the
scale of realistic calculations of physical hadron properties by a
decade or more!

\subsection{Chiral Loops and Non-Analyticity}
We have already seen that spontaneous chiral symmetry breaking in QCD
requires the existence of Goldstone bosons whose masses vanish in the
limit of zero quark mass (the chiral limit).
As a corollary to this, there must be
contributions to hadron properties from Goldstone boson loops.  These
loops have the unique property that they give rise to terms in an
expansion of most hadronic properties as a function of quark mass which
are not analytic \cite{CHIPT}. 
As a simple example, consider the nucleon mass.  The
most important chiral corrections to $M_N$ come from the processes
$N \ra N\pi \ra N$ ($\sigma_{NN}$) and $N \ra \Delta \pi \ra N$
($\sigma_{N \Delta}$). We write 
$M_N = M_N^{\rm bare} + \sigma_{NN} + \sigma_{N \Delta}$.
In the heavy baryon limit one has
\be
\sigma_{NN} = - \frac{3 g_A^2}{16 \pi^2 f_\pi^2}
\int_0^\infty dk \frac{k^4 u^2(k)}{k^2 + m_\pi^2}.
\label{eq:3A}
\ee
Here $u(k)$ is a natural high momentum cut-off which is the Fourier
transform of the source of the pion field (e.g. in the 
CBM it is $3 j_1(kR)/kR$, with $R$ the bag radius \cite{CBM}). From
the point of view of PCAC it is natural to identify $u(k)$ with the
axial
form-factor of the
nucleon, a dipole with mass parameter
$1.02 \pm 0.08$GeV \cite{Thomas:2001kw}.

Totally independent of the form chosen for the ultra-violet cut-off, one
finds that $\sigma_{NN}$ is a non-analytic function of the quark mass.
The non-analytic piece of $\sigma_{NN}$ is independent
of the form factor and gives
\be
\sigma_{NN}^{LNA} = - \frac{3 g_A^2}{32 \pi f_\pi^2} m_\pi^3
\sim \bar{m}^{\frac{3}{2}}.
\label{eq:4A}
\ee
This has a branch point, as a function of $\bar{m}$, starting at
$\bar{m} = 0$. Such terms can only arise from Goldstone boson loops.

It is natural to ask how significant this non-analytic behaviour
is in practice.  If
the pion mass is given in GeV, $\sigma_{NN}^{LNA} = -5.6 m_\pi^3$
and at the physical pion mass it is just -17 MeV.
However, at only three times the physical pion mass, $m_\pi = 420$MeV,
it is -460MeV -- half the mass of the nucleon. 
If one's aim is to extract
physical nucleon properties from lattice QCD calculations this is
extremely important.  As we explained earlier,
the most sophisticated lattice calculations with
dynamical fermions are only just becoming feasible at such low masses
and to connect to the physical world one must extrapolate from
$m_\pi \sim 500$MeV to $m_\pi = 140$MeV.
Clearly one must have control of the chiral behaviour.
{}Figure \ref{fig:FIG3} shows recent
lattice calculations of $M_N$ as a function of
$m_\pi^2$ from CP-PACS and UKQCD \cite{Aoki:1999ff,Allton:1999gi}.
The dashed line indicates a fit which
naively respects the presence of a LNA term,
\be
M_N = \alpha + \beta m_\pi^2 + \gamma m_\pi^3,
\label{eq:5}
\ee
with $\alpha, \beta$ and $\gamma$  fitted to
the data.  While this gives a very good fit to the data, the chiral
coefficient $\gamma$ is only -0.761, compared with the value -5.60
required by chiral
symmetry.  If one insists that $\gamma$ be 
consistent with QCD the best fit 
one can obtain with this form is the dash-dot curve.  This is clearly
unacceptable.
\begin{figure}[ht]
\centering{\
\epsfig{file=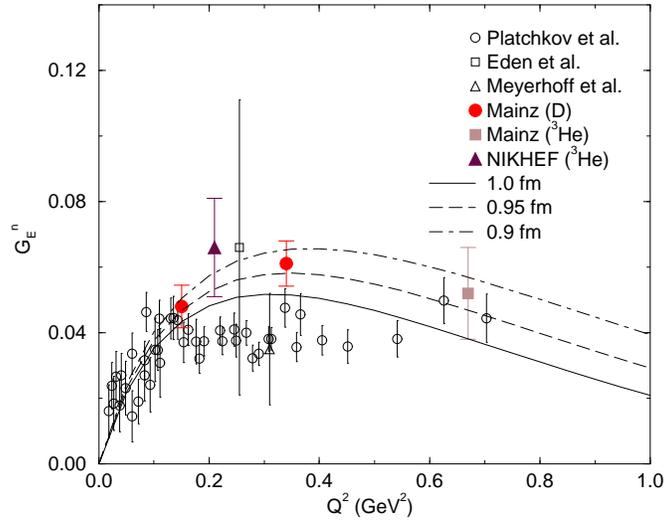,height=8cm}
\caption{Recent data for the neutron electric form factor
in comparison with CBM calculations for a
confining radius around 0.95fm -- from Ref.\ \protect\cite{Lu:2001yc}.}
\label{fig:neutron}}
\end{figure}

An alternative suggested recently by
Leinweber et al. \cite{Leinweber:2000ig}, which also
involves just three parameters, is to evaluate $\sigma_{NN}$ and
$\sigma_{N\Delta}$ with the same
ultra-violet form factor, with mass parameter $\Lambda$, and to fit
$M_N$ as
\be
M_N = \alpha + \beta m_\pi^2 + \sigma_{NN}(m_\pi,\Lambda) +
\sigma_{N\Delta}(m_\pi,\Lambda), 
\label{eq:6}
\ee
by adjusting $\alpha, \beta$ and $\Lambda$ to fit the data.
Using a sharp cut-off ($u(k) = \theta(\Lambda - k)$) these
authors were able to obtain
analytic expressions for $\sigma_{NN}$ and $\sigma_{N\Delta}$
which reveal the correct LNA
behaviour -- and next to leading (NLNA) in the $\Delta \pi$ case,
$\sigma_{N\Delta}^{NLNA} \sim
m_\pi^4 \ln m_\pi$.
These expressions also reveal a branch point at $m_\pi = M_\Delta -
M_N$,
which is important if one is extrapolating from large values of $m_\pi$
to the physical value.  The solid curve in Fig. \ref{fig:FIG3}
is a two parameter fit
to the lattice data using Eq.(\ref{eq:6}), but fixing $\Lambda$
at a value suggested by CBM simulations to be
equivalent to the prefered 1 GeV dipole. A small increase in $\Lambda$
is necessary to fit the lowest mass data point (at $m_\pi^2 \sim
0.1$ GeV$^2$) well, but clearly one can describe the data very 
satisfactorily while
preserving the exact LNA and NLNA behaviour of QCD.

The analysis of the lattice data for $M_N$, incorporating the correct
non-analytic behaviour, also yields important new information concerning
the sigma commutator of the nucleon:
\be
\sigma_N = \frac{1}{3} \langle N| [Q_{i 5},[Q_{i 5},H_{QCD}]] |N\rangle
= \langle N| \bar{m} (\bar{u} u + \bar{d} d) |N\rangle,
\label{eq:8}
\ee
which is a direct measure of chiral SU(2) symmetry breaking in QCD.
The widely accepted experimental value is
$45 \pm 8$MeV \cite{Gasser:1991ce}, although there
are recent suggestions that it might be as much as 20 MeV
larger \cite{Knecht:1999dp}.
Using the Feynman-Hellmann theorem one can also write
\be
\sigma_N = \bar{m} \frac{\partial M_N}{\partial \bar{m}}
= m_\pi^2 \frac{\partial M_N}{\partial m_{\pi}^2},
\label{eq:9}
\ee
at the physical pion mass.
Historically, lattice calculations have evaluated
$<N| (\bar{u} u + \bar{d} d) |N>$ at large quark mass
and extrapolated this
scale dependent quantity to the ``physical'' quark mass, which had to
be determined in a separate calculation.  The latest result with
dynamical fermions, $\sigma_N = 18 \pm 5$ MeV \cite{Gusken:1999wy},
illustrates how difficult this procedure is. On the other hand, if one
has a fit to $M_N$ as a function of $m_\pi$ which is
consistent with chiral symmetry, one can evaluate $\sigma_N$
directly using Eq.(\ref{eq:9}). Using Eq.(\ref{eq:6}) with a sharp
cut-off yields $\sigma_N \sim 55$ MeV, while a dipole form gives
$\sigma_N \sim 45$ MeV \cite{Wright:2000gg}. The residual model dependence can
only be
removed by more accurate lattice data at low $m_\pi^2$. Nevertheless,
the result $\sigma_N \in (45,55)$ MeV is in very good agreement with the
data.  In
contrast, the simple cubic fit, with $\gamma$ inconsistent with chiral
constraints, gives $ \sim 30$ MeV. Until the experimental situation
regarding $\sigma_N$ improves, it is not possible to draw definite
conclusions regarding the strangeness content of the
nucleon from this
analysis, but the fact that two-flavour QCD reproduces the current
prefered value should certainly stimulate more work.

%
\begin{figure}[hb]
\centering{\
\epsfig{file=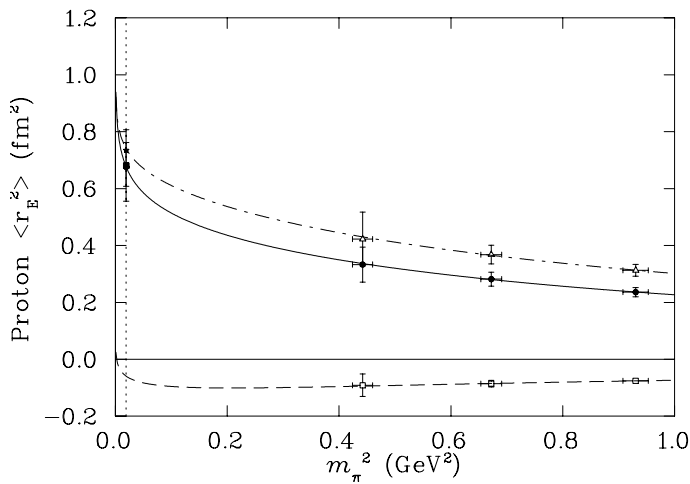, height=9cm, angle=90}
\caption{Fits to lattice results for the squared electric charge radius
of the proton -- from Ref. \protect\cite{Hackett-Jones:2000js}. Fits
to the contributions from individual quark flavors are also
shown: the $u$-quark sector results are indicated by open triangles
and the $d$-quark sector results by open squares. Physical values
predicted by the fits are indicated at the physical pion mass, where
the full circle denotes the result predicted from the first
extrapolation procedure and the full square denotes the baryon radius
reconstructed from the individual quark flavor extrapolations. (N.B.
The latter values are actually so close as to be
indistinguishable on the graph.) The
experimental value is denoted by an asterisk.}
\label{fig:prot}}
\end{figure}
%
\subsection{Electromagnetic Properties of Hadrons}
It is a completely general consequence of quantum mechanics that the
long-range charge structure of the proton comes from its $\pi^+$ cloud
($p \ra n \pi^+$),
while for the neutron it comes from its $\pi^-$ cloud ($n \ra p \pi^-$).
However it is not
often realized that the LNA contribution to the nucleon charge radius
goes like $\ln m_\pi$ and diverges as $\bar{m} \ra 0$ \cite{Leinweber:1993hj}.
This can never be described by a
constituent quark model. Figure \ref{fig:neutron}
shows the latest data from Mainz and Nikhef
for the neutron electric form factor, in comparison with CBM
calculations for a confinement radius between 0.9 and 1.0 fm. The
long-range $\pi^-$ tail of the neutron plays a crucial role 
\cite{Lu:2001yc,Lu:1998sd}.

While there is only limited (and indeed quite old) lattice data for
hadron charge radii, recent experimental progress in the determination
of hyperon charge radii has led us to examine the extrapolation
procedure for obtaining charge data from the lattice simulations
\cite{Hackett-Jones:2000js}. Figure \ref{fig:prot} 
shows the extrapolation of the
lattice data for the charge radius of the proton. Clearly the
agreement with experiment is much better once the logarithm required by
chiral symmetry is correctly included, than if, for example, one simply
makes a linear extrapolation in the quark mass (or $m_\pi^2$).
{}Full details of the results for all the octet baryons may be found in
Ref. \cite{Hackett-Jones:2000js}.
\begin{figure}[htb]
\centering{\
\epsfig{file=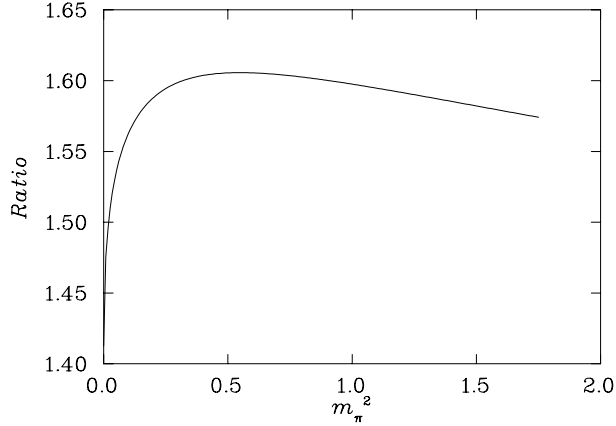, height=8cm, angle=90}
\caption{Ratio of the (modulus of the) 
proton to neutron magnetic moments as a function
of $m_\pi^2$, obtained from a simple Pad\'e approximant which includes
the correct LNA behaviour
-- from Ref. \protect \cite{Leinweber:2001ui}. 
}
\label{fig:ratio}}
\end{figure}

The situation for baryon magnetic moments is also very interesting.
The LNA contribution in this case arises from the diagram where
the photon couples to the pion loop.  As this
involves two pion propagators the expansion of the proton and neutron
moments is:
\be
\mu^{p(n)} = \mu^{p(n)}_0 \mp \alpha m_\pi + {\cal O}(m_\pi^2).
\label{eq:10}
\ee
Here $\mu^{p(n)}_0$ is the value in the chiral limit and the
linear term in $m_\pi$ is proportional to $\bar{m}^{\frac{1}{2}}$,
a branch point at $\bar{m} = 0$.  The
coefficient of the LNA term is $\alpha = 4.4 \mu_N $GeV$^{-1}$.
At the physical pion mass this LNA
contribution is $0.6\mu_N$, which is almost a third of the neutron
magnetic moment.

Just as for $M_N$, the chiral behaviour of $\mu^{p(n)}$ is vital to a
correct
extrapolation of lattice data. One can obtain a very satisfactory fit to
some rather old data, which happens to be the best available,
using the simple Pad\'e \cite{Leinweber:1999ej}:
\be
\mu^{p(n)} = \frac{\mu^{p(n)}_0}{1 \pm \frac{\alpha}{\mu^{p(n)}_0} m_\pi
+
\beta m_\pi^2}
\label{eq:11}
\ee
Existing lattice data can only determine two parameters and Eq.(\ref{eq:11})
has just two free parameters while guaranteeing the correct LNA
behaviour as $m_\pi \ra 0$ {\bf and} the correct behaviour of HQET
at large $m_\pi^2$.  The
extrapolated values of $\mu^p$ and $\mu^n$
at the physical pion mass, $2.85 \pm 0.22 \mu_N$ and $-1.90 \pm 0.15
\mu_N$ are currently the best estimates from non-perturbative QCD
\cite{Leinweber:1999ej}. For the application
of similar ideas to other members of the
nucleon octet we refer to Ref. \cite{Hackett-Jones:2000qk}, and for the
strangeness magnetic moment of the
nucleon we refer to Ref. \cite{Leinweber:2000nf}.
The last example is another case where tremendous improvements in the
experimental capabilities,  
specifically the accurate measurement of
parity violation in $ep$ scattering 
\cite{Spayde:2000qg,Kumar:2000eq,Souder:2000vw},
is giving us vital information on
hadron structure.

In closing, we note that from the point of view of the naive quark model it is
interesting to plot the ratio of the proton to neutron magnetic moments
as a function of $m_\pi^2$. The closeness of the experimental value to
-3/2 is usually taken as a major success. However, we see from Fig.
\ref{fig:ratio} that it is in fact a matter of luck
\cite{Leinweber:2001ui}. We stress that the
large slope of the ratio near $m_\pi^2 = 0$ is {\em model independent}.

\subsection{Moments of Structure Functions}
The moments of the parton distributions measured in lepton-nucleon
deep inelastic scattering \cite{Thomas:2001kw} are defined as:
\begin{eqnarray}
\langle x^n \rangle_q
&=& \int_0^1 dx\ x^n\
\left( q(x,Q^2) + (-1)^{n+1} \bar q(x,Q^2) \right)\ ,
\label{eq:mom}
\end{eqnarray}
where the quark distribution $q(x,Q^2)$ is a function of the
Bj\"orken
scaling variable $x$ (at momentum scale $Q^2$).  Then the operator
product expansion relates these moments to the
{}forward nucleon matrix elements of certain local twist-2 operators
which can be accessed in lattice simulations.

\begin{figure}
\centering{\
\hspace*{.7cm}\epsfig{file=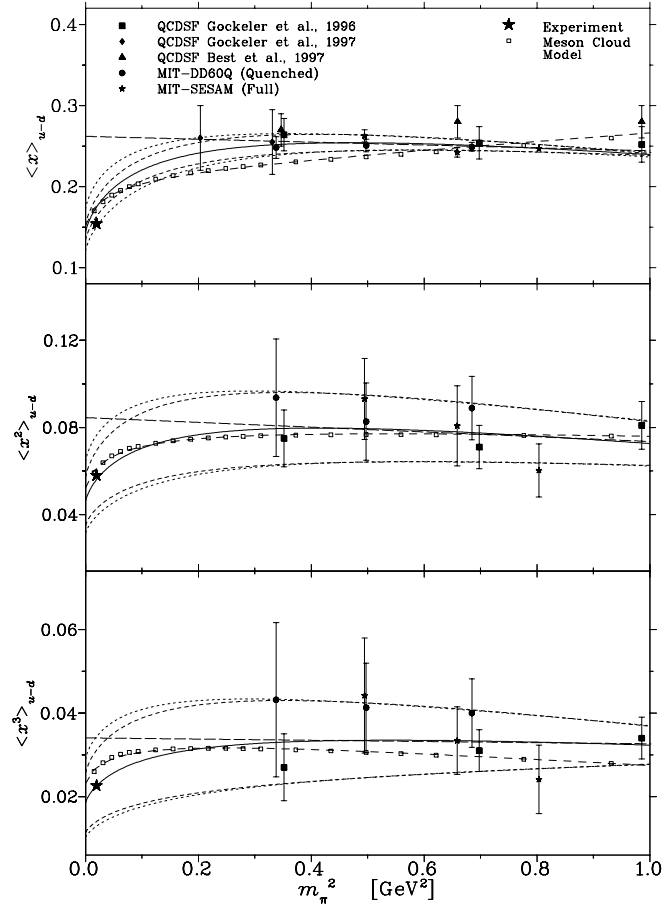,height=12cm}
\caption{Moments of the $u - d$ quark distribution from various lattice
simulations.
The straight (long-dashed) lines are linear fits
to this data, while the curves have the correct LNA behaviour in
the chiral limit -- see the text for details.
The small squares are the results of the meson cloud model
and the dashed curve through them
best fits using Eq.~(\protect\ref{eq:fit}).  The star represents the
phenomenological values taken from NLO fits in
the $\overline{\rm MS}$ scheme.}
\label{mainfig}}
\end{figure}

Early calculations of moments of structure functions within lattice
QCD were performed by Martinelli and Sachrajda
\cite{Martinelli:1987zd}.  However, the
more recent data used in this analysis are taken from the
QCDSF \cite{Gockeler:1997jk} and MIT 
\cite{Dolgov:2001ca} groups and shown in Fig.~\ref{mainfig} for the
$n=1$, 2 and 3 moments of the $u - d$ difference
at NLO in the $\overline{\rm MS}$ scheme.
These calculations have been performed for both full
and quenched QCD using a
variety of quark actions and for quark masses, $\bar m$, ranging from 50 to
190~MeV.

To compare the lattice results with the experimentally measured
moments, one must extrapolate the data from the lowest quark mass used
($\sim 50$~MeV) to the physical value ($\sim$ 5--6~MeV).
Naively this is done by assuming that the moments depend linearly on the
quark mass.  However, as shown in Fig.~\ref{mainfig} (long dashed
lines), a linear extrapolation of the world lattice data for the $u-d$
moments typically overestimates the experimental values by 50\%.  This
suggests that important physics is still being omitted from the
lattice calculations and their extrapolations.

Here, as for all other hadron properties, a linear extrapolation
in $\bar m \sim m_\pi^2$ must fail as it omits crucial nonanalytic
structure associated with chiral symmetry breaking.  
The studies of the chiral extrapolation of lattice data for
hadron masses, magnetic moments and charge
radii, which we have just reviewed, 
have shown that for quark masses above 50--60~MeV,
hadron properties behave very much as one would expect in a
constituent quark model, with relatively smooth behaviour as a
function of the quark mass.  However, for $\bar m < 50$~MeV one typically
finds rapid, nonlinear variation arising from the nonanalytic behaviour
of Goldstone boson loops \cite{CHIPT}.

In general, contributions to the physical properties of hadrons from
intermediate states involving the surrounding meson cloud give rise
to unique terms which are nonanalytic in the quark mass.  These stem
from the infra-red behaviour of the chiral loops and are {\em model
independent}.  The leading nonanalytic (LNA) term for the $u$ and
$d$ distributions in the physical nucleon arises from the single pion
loop dressing of the bare nucleon and has been
shown \cite{Thomas:2000ny,LNA_DIS} to
behave as:
\begin{equation}
\langle x^n \rangle_q^{\rm LNA}
\sim m_\pi^2 \log m_\pi\ .
\label{eq:lna}
\end{equation}

Experience with the chiral behaviour of masses and magnetic moments
shows that the LNA terms alone are not sufficient to describe lattice
data for $m_\pi > 200$~MeV.  Thus, in order to fit the 
lattice data at larger $m_\pi$, while preserving the correct chiral
behaviour of moments as $m_\pi \to 0$, a low order, analytic expansion
in $m_\pi^2$ is also included in the extrapolation and the moments of
$u-d$ are fitted with the form \cite{Detmold:2001jb}:
\begin{equation}
\label{eq:fit}
\langle x^n \rangle_{u-d}
=\ a_n\ +\ b_n\ m_\pi^2\
+\ a_n\ c_{\rm LNA}\ m_\pi^2
\ln \left( \frac{m_\pi^2}{m_\pi^2 + \mu^2} \right)\ ,
\end{equation}
where the coefficient $c_{\rm LNA} = -(3 g_A^2+1)/(4\pi f_\pi)^2$
\cite{LNA_DIS}. The parameters $a_n$,
$b_n$ and $\mu$ are {\em a priori} undetermined.  The mass $\mu$
determines the scale above which pion loops no longer yield rapid
variation and corresponds to the upper limit of the momentum
integration if one applies a sharp cut-off in the pion loop integral.
Multi-meson loops and other contributions cannot give rise to the LNA
behaviour in Eq.~(\ref{eq:lna}) and thus near the chiral limit
Eq.~(\ref{eq:fit}) is the most general form for moments of the PDFs at
${\cal O}(m_\pi^2)$ which is consistent with chiral symmetry.

Having motivated the functional form of the extrapolation formula, we
now apply Eq.~(\ref{eq:fit}) to the lattice data.  Unfortunately, data
are not yet available at quark masses low enough to allow a reliable
determination of the mass parameter $\mu$.
Consequently, for the central curve in each panel of Fig.~\ref{mainfig}
the value that is most consistent with all experimental moments was
chosen,
$\mu=550$~MeV. With $\mu$ thus fixed, the results of the best $\chi^2$
fit (for parameters $a_n$ and $b_n$) to the lattice data for each
moment are given by the central solid lines.  To
estimate the error in the extrapolated value (for a fixed $\mu$), we
also fit to the extrema of the error bars on the data as is shown in
Fig.~\ref{mainfig} by the inner envelopes around these curves.

Experience with other hadronic properties,
such as magnetic moments and
masses, suggests that the switch (as a function of current quark
mass) from smooth and constituent quark-like behaviour (slowly
varying with respect to the current quark mass) to rapidly variation
(dominated by Goldstone boson loops) happens for $m_\pi \sim $ 500--600
MeV. This is very close to the preferred value of $\mu$ found here
and the similarity of these scales
for the various observables simply reflects the common scale at which
the Compton wavelength of the pion becomes comparable to the size of
the bare hadron. 
Nevertheless, we stress that $\mu$ is not yet determined by the
lattice data and it is indeed possible
to consistently fit both the lattice
data and the experimental values with $\mu$ ranging from
400~MeV to 700~MeV. This dependence on $\mu$ is illustrated in
Fig.~\ref{mainfig} by the difference between the inner and outer
envelopes on the fits.  The former are the best fits to the lower
(upper) limits of the error bars, while the latter use $\mu=450$
(650)~MeV instead of the central value of $\mu=550$~MeV. Data at
smaller quark masses, ideally $m_\pi^2 \sim$ 0.05--0.10 GeV$^2$,
are therefore crucial to constrain this parameter
and perform an accurate extrapolation.
\subsection{Baryon Spectroscopy}
The study of the baryon spectrum is a venerable art \cite{Thomas:2001kw}. 
However, the lack
of suitable experimental facilities has meant that there has been
insufficient data to provide definitive tests for the many theoretical
models constructed over the past 30 years.  The availability of high
intensity, high duty factor electron accelerators, complemented with
multi-particle detectors, means that this situation is changing
dramatically.  Amongst the open questions to be addressed initially one
might ask:
\begin{itemize}

\item What is the Roper resonance ($R(1440)$)?
In a naive quark model it would be a $2 \hbar \omega$ 
excitation of the nucleon, yet it lies below the $1 \hbar \omega$
negative parity states. Is it a breathing mode \cite{Guichon:1985nw} 
or a channel coupling effect \cite{Speth:2000zf}?

\item Is the $\Lambda(1405)$ a $\bar K N$ bound state,
as suggested originally by Dalitz and Tuan? Is it a result of the
coupling of the $\Sigma \pi$--$\bar K N$ channels, taking into
account the extremely attractive $\Sigma \pi$ interaction near
threshold \cite{Veit:1985jr,Kaiser:1995eg}?

\item Do the missing states, predicted by the quark model but not yet
seen experimentally, really exist?

\item Are there some states which are not described by the quark model
at all, but simply a consequence of very strong rescattering?
\end{itemize}

We may expect a great deal of experimental insight into these questions
in the next few years, from JLab, Mainz and MIT-Bates.  At the same
time, there are also some exciting developments on the theoretical
side.  Until now we have had an over-abundance of models, more or less
motivated by QCD, with no rigorous way to choose between them.  The
recent progress in lattice QCD will also have a dramatic impact here.
Pioneering work on the $1 \hbar \omega$ and $2 \hbar \omega$ 
excited states of the nucleon by
Leinweber \cite{Leinweber:1995nm}, and by 
Leinweber and Lee \cite{Lee:1999cx}, is now being developed by at least
three groups: at BNL \cite{Sasaki:2001nf}, JLab--QCDSF--UKQCD 
\cite{Gockeler:2001db} and CSSM \cite{Zanotti}.

The masses of the $N'(1/2^+), N^*(1/2^-), \Delta^*(3/2^-)$ and even the
strange excited states are now being studied in detail  with a variety
of non-perturbative improvements of the action as well as with domain
wall fermions, to improve the chiral properties of the calculation.
Again the
computer limitations mean that all calculation so far have been quenched
and also restricted to relatively large quark mass.  Nevertheless some
features are already apparent.  The data looks very much as one would
expect in a constituent quark model.  The masses come in the order
expected in a simple oscillator picture with $2 \hbar \omega$ (positive parity)
excitations well above the $1 \hbar \omega$ states.
Consistent with our earlier summary of the large mass region ($\bar m$
above 60MeV), the masses are linear
functions of the quark mass.  It is clearly vital to extend the
calculations to quark masses that are as low as possible and to explore
the chiral constraints on the extrapolations of these excited states.
There will be tremendous progress in this field in the next five years.

\section{Nuclear Systems}
\begin{figure}
\centering{\
\epsfig{file=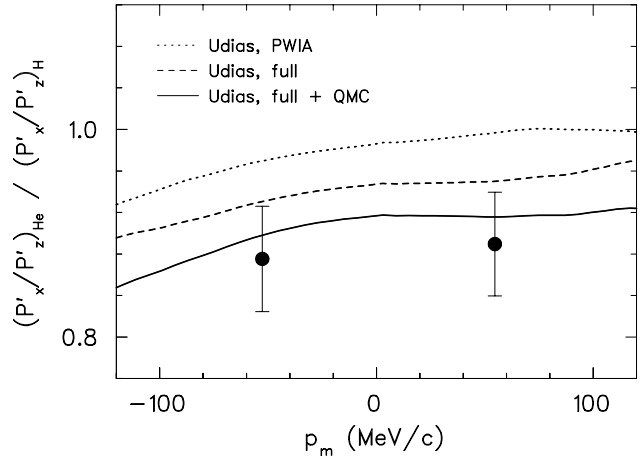, angle=90, height=6cm}
\caption{Comparison between the ratio of polarised electron
quasi-elastic scattering cross sections in $^4$He and on the free proton.
The measured ratio is proportional to the ratio of the proton electric to
magnetic form factors of a proton bound in $^4$He to the ratio for a
free proton if the other nuclear effects are correctly handled. The data
favour earlier calculations within QMC which predict a change in the
bound nucleon form factors. The figure is taken from
Ref. \protect \cite{Dieterich:2001mu}.}
\label{fig:bound}}
\end{figure}
The traditional view of nuclei is as a system of unperturbable
(point-like) protons and neutrons.  In a mean-field treatment they move
in a self-consistent binding potential.  Within quantum hadrodynamics
(QHD) the Lorentz character of the mean-field is taken seriously, with a
strong scalar attraction of order 300-400 MeV at nuclear matter density
($\rho_0$) and an almost equally strong vector repulsion
\cite{Furnstahl:2000in}.  The
intermediate range scalar attraction is readily understood in terms of
the two-pion-exchange interaction built into the Paris or Bonn
potentials, while the vector repulsion involves $\omega$--exchange
(possibly with short distance quark exchange).  From the point of view
of effective field theory this conventional approach is perfectly
satisfying and one seeks to determine the additional terms in a complete
effective Lagrangian which reproduce all nuclear phenomena
\cite{Serot:2000pv}.  This will
involve non-linear meson-baryon couplings and density dependent
effective interactions.

On the other hand, believing that we have a complete theory of the
strong interactions, namely QCD, it is also fascinating to ask what
nuclear phenomena we can understand at that more fundamental level.
Indeed, including the transition to a quark-gluon plasma, it is
essential to tune one's theoretical ideas at the lower densities of
normal nuclei.

If one starts with the aim of understanding nuclear structure in terms
of QCD it is fundamental that the protons and neutrons are far from
elementary.  Indeed, they are quite large composite systems of quarks
and gluons.  It has been recognized for more than 20 years that the
typical nearest neighbour separation at $\rho_0$ is very close to twice
the radius of the nucleon.  Furthermore, the typical mean scalar field
noted earlier is exactly the same size as the energy required to excite
a nucleon (300MeV to the $\Delta$ or 500 MeV to the $R(1440)$).  {\bf
{}Far from being a surprise that nucleon structure might play a role in
nuclear structure, it is difficult to see how it could fail to be
significant!  One of the central aims of modern nuclear physics must be
to investigate this role both theoretically and experimentally.}

The quest for changes in the structure of hadrons in medium began in the
80's, with considerable attention to the problem being generated by the
idea of Brown-Rho scaling \cite{Brown:1991kk}. 
At about the same time, Guichon constructed
a simple generalization of QHD, in which the point-like nucleon was
replaced by an MIT bag and the $\sigma$ and $\omega$ mesons coupled to
the confined quarks \cite{Guichon:1988jp}. 
A self-consistent solution of this problem within
mean-field approximation led to an astonishing result; the response of
the wave function of the confined quarks
to the scalar field led to a natural saturation mechanism.  That is,
rather than the $\sigma-N$ coupling being a constant it becomes a
monotonically decreasing function of the applied scalar field, $g_\sigma
(\sigma)$.  Indeed, because of the low mass of the confined quarks this
mechanism is far more effective than the decrease of $\bar \psi_N \psi_
N$ in QHD, and the mean scalar field at $\rho_0$ is only 50\% or so of
that in QHD.
Technically, the expression for the energy of nuclear matter in the
Guichon model is identical to that in QHD.  The only place that the
internal structure of the nucleon enters is in the equation for the mean
scalar field, where the $\sigma-N$ coupling constant is replaced by
$\partial M_N^*/ \partial \bar \sigma$.  The latter {\bf does} depend on
the structure of the nucleon -- it is essentially its scalar response.
Now, this is a quantity which could be calculated on the lattice once
one has sufficient control to go to near realistic quark mass.  In this
way one could actually begin to investigate nuclear saturation with
input from QCD itself!

This model, which is generally known as QMC (the quark-meson coupling
model), has been extensively developed by Guichon, Saito, Tsushima,
Blunden, Miller, Jennings and many others 
\cite{Guichon:1996ue,Saito:1997yb,Blunden:1996kc,Jin:1996qa}.  
Exchange contributions have been
considered \cite{Krein:1999vc}
and eventually one must go to a more sophisticated
theoretical treatment than simple mean field theory.  Nevertheless, the
model gives interesting guidance on the 
modification of hadron masses \cite{Tsushima:1998qw}
and reaction cross sections inside nuclear 
matter \cite{Tsushima:2001re}.  For instance, it
provides an interesting alternative to the naive explanation
of $J/ \Psi$ suppression in relativistic heavy ion 
collisions in terms of a quark-gluon plasma \cite{Sibirtsev:2000jr}.

{}From the point of view of nuclear structure the most interesting
development is the application to finite nuclei.  One can show that, at
least at mean-field level, Born-Oppenheimer approximation in which the
quark motion adjusts to the local mean-scalar field at any given place,
should be good to about 3\% for normal nuclei.  Then one can derive a
nuclear shell model in which the nucleon self-consistently adjusts to
the local mean scalar field in each single 
particle orbital \cite{Guichon:1996ue}.  On the
other hand, this derivation implies a {\bf deep conceptual change in our
understanding}.  What occupies the shell model orbits are not nucleons
but nucleon-like quasi-particles.  These will have different masses,
magnetic moments, charge radii and so on from those of free nucleons.
{}From this point of view is less remarkable that bound nucleons should
have different properties from free nucleons than that such changes have
proven so difficult to establish.  This is why the nuclear 
EMC effect \cite{Aubert:1983xm,Geesaman:1995yd},
which is still only partially understood, was so important.

In terms of a fundamental theory of nuclear structure there
can be few more important challenges than establishing the change in the
properties of a bound nucleon.  In this respect, a recent experimental
result from Mainz is potentially extremely important
\cite{Dieterich:2001mu}.  This group used
the same triple scattering technique which was so effectively used to
determine $G_E/G_M$ for the free proton at JLab, to measure $G_E/G_M$
{}for a proton bound in $^4$He using the $(\vec{e}, e' \vec{p})$
reaction. The 
change in the ratio of these form factors for the bound nucleon had been
studied in detail within QMC \cite{Lu:1999tn,Thomas:1998eu}
and, as shown in Fig.~\ref{fig:bound}, the experimental
results support such a modification.  Clearly the statistical
significance of the effect is not yet great.  On the other hand, careful
theoretical study of the effects of distortion, spin-orbit forces and meson
exchange suggests that this particular ratio is extremely
insensitive to such corrections.  This measurement is crucial in that it
is really the first clear indication of a change in the structure of a
bound nucleon.  It will stimulate a great deal more work!

\section{Conclusion}
This is indeed an exciting period in the development of hadron physics.
We have seen that developments in lattice QCD, especially more powerful
computers and improved chiral extrapolations, should finally allow the
computation of accurate hadron properties within full QCD,
{\em at the physical quark
masses}, within the next five years.
We can also expect new insights into the structure of the QCD vacuum,
the nature of confinement and the mechanism for spontaneous chiral
symmetry breaking. Studies of hadron spectroscopy on the lattice will
complement important new experimental studies and improved quark
models. 

We have seen that from the point of view of QCD, it is vital to
understand the changes in hadron properties that occur as a function of
density as well as temperature. In this sense finite nuclei provide a
crucial testing ground for ideas that will eventually be applied at much
more extreme conditions. Crucial experiments involve the possible
binding of $\omega$, $\eta$ and even charmed mesons in finite nuclei, as
well as the changes in the form factors of bound nucleons expected
within QCD and predicted within various QCD motivated models. We briefly
reviewed an experiment that gives a tantalizing hint of such a change
and look forward to the many further tests of these ideas that will
follow in the next few years. 

\begin{theacknowledgments}
It is a pleasure to acknowledge the contributions from many
collaborators in the development of the ideas presented here. I would
especially like to thank my colleagues at CSSM and particularly Derek
Leinweber, Wally Melnitchouk and Tony Williams. This
work was supported by the Australian Research Council and the University
of Adelaide.
\end{theacknowledgments}


\bibliographystyle{aipproc}

\end{document}